\documentclass[12pt]{article}

\def\be{\begin{equation}}
\def\ee{\end{equation}}
\def\ben{\begin{displaymath}}
\def\een{\end{displaymath}}
\def\ba{\begin{array}{c}}
\def\ea{\end{array}}

\newcommand{\kt}{\rangle}
\newcommand{\br}{\langle}
\newcommand{\ed}{\end{document}}
\newcommand{\bbr}{\br\!\br}
\newcommand{\kkt}{\kt\!\kt}
\newcommand{\pbr}{\prec\!}
\newcommand{\pkt}{\!\succ\,}

\newcommand{\brz}{\{\!\{}
\newcommand{\ktz}{\}\!\}}

\begin{document}

\begin{center}

{\Large \bf

${\cal PT}-$symmetric Sturmians
  }

\vspace{9mm}

{Miloslav Znojil}

\vspace{9mm}

Nuclear Physics Institute ASCR, 250 68 \v{R}e\v{z}, Czech Republic

email: {znojil@ujf.cas.cz}


\end{center}


\section*{Abstract}

Sturmian bound states emerging at a fixed energy and numbered by a
complete set of real eigencouplings ${\lambda}_n$  are considered.
For Sturm-Schr\"{o}dinger equations which are manifestly
non-Hermitian we outline the way along which the correct
probabilistic interpretation of the system can constructively be
re-established via a new formula for the metric. ${\cal
PT}-$symmetrized Coulomb potential is chosen for illustration
purposes.

\newpage

\section{Introduction}

During the early developments of Quantum Mechanics,
Schr\"{o}dinger equation for the Coulomb bound-state problem
 \be
 -\frac{d^2}{dr^2}\,\psi_{n,\ell}^{({\lambda})}(r)
 +\frac{\ell(\ell+1)}{r^2}
  \,\psi_{n,\ell}^{({\lambda})}(r)-
 \frac{{\lambda}}{r}\,\psi_{n,\ell}^{({\lambda})}(r)=
 -\kappa^2_{n,\ell}\,\psi_{n,\ell}^{({\lambda})}(r)\,,
 \label{SEracoul}
 \ee
 \ben
 \ \ \ \ \ \ \ \ \ \ \ \
 \ \ \ \ \ \ \ \ \ \ \ \
 \ \ \ \ \ \ \ \ \ \ \ \
  \ \ \ \ \ \ n,\ell=0,1,\ldots\,,\ \ \ \ \ {\lambda} >0
 \een
played role in the successful quantitative description of
hydrogen-like atoms as well as in the various theoretical
considerations verifying, e.g., the possibility of the coexistence
of the discrete and continuous spectra. Although the purely
phenomenological appeal of the oversimplified model
(\ref{SEracoul}) has perceivably weakened with time, its
mathematical and methodical relevance seems to survive in full
strength. This is the reason why we choose here this example for
illustrative purposes. Before we formulate our main results, let
us recollect at least a part of the necessary terminology
concerning, first of all, the so called Sturmians and the so
called ${\cal PT}-$symmetry.

\subsection{Sturmians}

One of the most important reasons of the nondecreasing popularity
of the Coulombic quantum model (\ref{SEracoul}) lies in its exact
solvability in terms of Laguerre polynomials \cite{Fluegge}.
Recently, Kelbert et al \cite{Kelbert} emphasized that one of the
particularly important aspects of this solvability should be seen
in a certain symmetry between the roles played by the bound-state
energies $E= -\kappa^2$ and by the related strengths of the
Coulombic attraction or charge ${\lambda} >0$.

Due to such a symmetry one can rearrange eq.~(\ref{SEracoul}) in
such a way that the energy value $E= -\kappa^2$ is kept fixed
while the ``bound-state-supporting" eigencharges ${\lambda}>0$ are
allowed to vary,
 \be
 -\frac{d^2}{dr^2}\,\phi_{n,\ell}^{(\kappa)}(r)
 +\frac{\ell(\ell+1)}{r^2}
  \,\phi_{n,\ell}^{(\kappa)}(r)
 +\kappa^2\,\phi_{n,\ell}^{(\kappa)}(r)
 =
 {\lambda}_{n,\ell}\,W(r)
 \,\phi_{n,\ell}^{(\kappa)}(r)
  \,.
 \label{SEstcoul}
 \ee
The original nonconstant potential $W(r)=W^{(Coulombic)}(r)=1/r$
now plays the role of a non-trivial factor accompanying the {new}
eigenvalue ${\lambda}>0$. On the level of terminology, equation
(\ref{SEstcoul}) replaces the original bound states
$\psi_{n,\ell}^{({\lambda})}(r)$ by another set of the
$\kappa-$dependent bound states $\phi_{n,\ell}^{(\kappa)}(r)$
called {\em Sturmians}.

The overall theory and applications of Sturmian wave functions
have been reviewed by Rotenberg \cite{Rotenberg} who covered areas
ranging from the electron-hydrogen scattering to the interatomic
charge-transfer collisions and solutions of the Faddeev and
Born-Oppenheimer equations. In  mathematical physics the Coulombic
as well as non-Coulombic Sturmians find typical applications in
perturbation theory \cite{rnad}. Their use has also been reported
to improve the convergence of certain non-perturbative algorithms
and calculations \cite{Kelbert}. Besides the immediate
determination of the energy-dependent couplings the
phenomenological use of Sturmians involves constructions of the so
called quasi-exactly solvable quantum models where, typically,
people use $W(r)\sim r^M$ with a positive (half)integer
$M$~\cite{Voros} and where due attention must be paid to the
questions of completeness \cite{refe}. Last but not least, in
connection with the so called resonant internal boundary layers
certain exact Sturmians emerging at $W(r)=r^N$ with any $N = -1,
0, 1, 2, \ldots$ even proved helpful in the context of classical
physics \cite{BW}.

\subsection{${\cal PT}-$symmetry}

The recent growth of interest in  manifestly non-Hermitian quantum
Hamiltonians $H$ with real spectra \cite{Carl} has originally been
motivated by the Bessis' imaginary cubic example $H=\hat{p}^2+{\rm
i}\hat{x}^3 \neq H^\dagger$ and by its ``symmetry" with respect to
the antilinear product ${\cal PT}$ of the space and time
reflections \cite{BB}. In the framework of the resulting ${\cal
PT}-$symmetric Quantum Mechanics (PTSQM, \cite{Carl}) it has been
explained, in ref.~\cite{sgezou}, why the well known and much
appreciated exact solvability of the standard Hermitian version of
the Coulomb problem (\ref{SEracoul}) survives its non-Hermitian,
${\cal PT}-$symmetric deformation.

In what follows we intend to complement the ${\cal
PT}-$symmetrized version of eq.~(\ref{SEracoul}) (and/or of its
various generalizations) by a transfer of the key ideas and
methodical conclusions of refs.~\cite{Carl,sgezou} to the parallel
non-Hermitian Sturmian problems.

Returning once more to the guiding ${\cal PT}-$symmetrized Coulomb
bound-state example of ref.~\cite{sgezou}, let us remind the
readers that the real potential  $W^{(Coulombic)}(r)=1/r$ has been
replaced there by its purely imaginary alternative $W^{({\cal
PT})}(r)={\rm i}/r$. In the resulting eq.~(\ref{SEracoul}), viz.,
 \be
 -\frac{d^2}{dr^2}\,\tilde{\psi}_{n,\ell}^{({\lambda})}(r)
 +\frac{\ell(\ell+1)}{r^2}
  \,\tilde{\psi}_{n,\ell}^{({\lambda})}(r)-{\rm i}\,
 \frac{{\lambda}}{r}\,\tilde{\psi}_{n,\ell}^{({\lambda})}(r)=
 -\kappa^2_{n,\ell}\,\tilde{\psi}_{n,\ell}^{({\lambda})}(r)\,
 \label{SEracoulPT}
 \ee
the manifest loss of the Hermiticity of the model has partially
been compensated by its ${\cal PT}-$symmetrization. This has been
achieved by the replacement of the most usual real half-line of
coordinates $r \in (0,\infty)$  by a left-right symmetric complex
parabola, typically, of the one-parametric form using a new real
coordinate $x$,
 \ben
 r=r(x) = 2\,x + {\rm i}\,(x^2-1)\,,
 \ \ \ \ \ \ \ \ x \in (-\infty,\infty)\,.
 \een
The original concept of the partial waves had to be abandoned so
that the kinematical centrifugal force $\ell(\ell+1)/r^2$ acquired
an immediate dynamical meaning and the choice of the parameter
$\ell$ was not restricted to (half)integers, anymore.

The main, slightly unexpected result of ref.~\cite{sgezou} was
that the spectrum of the bound-state energies was positive,
 \ben
 E_{(n,q)}=   \frac{{\lambda}^2}{ (2n+1-q-2\,q\,\ell)^2} \,,
 \ \ \ \ \ \ell \in I\!\!R\,,
 \ \ \ \ \
 n = 0, 1, \ldots \,
 \label{energs}
 \een
and that there emerged a new quantum number $q=\pm 1$ called
``quasi-parity". In a way encouraged by such an amazing structure
of levels offered by the model, our present letter will try to
extend the perspective of ref. \cite{sgezou} towards an analysis
of the family of its Sturmian eigenstates. In this spirit we shall
parallel the above-described transition from the bound states of
eq.~(\ref{SEracoul}) to the Sturmians of eq.~(\ref{SEstcoul})
under the additional assumption that one relaxes the Hermiticity
of the equations.

\subsection{${\cal PT}-$symmetric Sturmians}

Our transition to the ${\cal PT}-$symmetric Sturmian problem will
start directly from the ${\cal PT}-$symmetric Schr\"{o}dinger
bound-state equation~(\ref{SEracoulPT}). In the way guided by the
Hermitian case we shall interchange the role of the energy and
coupling, $\kappa^2 \leftrightarrow {\lambda}$. This leads to the
generalized eigenvalue problem
 \be
 -\frac{d^2}{dr^2}\,\tilde{\phi}_{n,\ell}^{(\kappa)}(r)
 +\frac{\ell(\ell+1)}{r^2} \,\tilde{\phi}_{n,\ell}^{(\kappa)}(r)+
 \kappa^2\,\tilde{\phi}_{n,\ell}^{(\kappa)}(r)
 =
 {\lambda}_{n,\ell}\,W^{({\cal PT})}(r)\,
 \tilde{\phi}_{n,\ell}^{(\kappa)}(r)
 \,
  \label{SEstcoulPT}
 \ee
where, as we already specified earlier, we may choose $W^{({\cal
PT})}(r)={\rm i}/r$ for the sake of definiteness.

Equation (\ref{SEstcoulPT}) and its physical meaning will be the
main targets of our present analysis. We shall identify the key
problem as lying in the correct specification of the concept of
observables. In the language of mathematics this means that we
shall describe a constructive recipe giving the metric operator
$\Theta$ in the Hilbert space ${\cal H}^{(physical)}$ of states of
our model. This will guarantee the existence as well as the form
of the standard probabilistic interpretation of its predictions.

In section \ref{druha} we shall start our considerations by
briefly reviewing the well known basic theory and methods in the
simpler, non-Sturmian case where $W=I$. We shall emphasize that
the complete solvability/solution of the underlying
Schr\"{o}dinger equation in an auxiliary, {\em unphysical} but
technically {\em strongly preferred} Hilbert space ${\cal
H}^{(user-friendly)}$ is usually assumed \cite{Carl,Geyer}. This
key methodical assumption enables us to make use of the {\em
general and explicit} infinite-series formula for the necessary
metric operator $\Theta$ at $W=I$ which is attributed, usually, to
Mostafazadeh \cite{Alib}.

In the next section \ref{tretia} the Sturmians with $W \neq I$
will be addressed and we shall describe the related generalization
of the PTSQM formalism. We shall slightly reorder the flow of the
arguments of section \ref{druha} and, after the introduction of
this generalized eigenvalue problem in the space ${\cal
H}^{(user-friendly)}$ we shall jump immediately to the abstract
analysis performed in another, ``inaccessible" Hilbert space
${\cal H}^{(third)}$ (cf. section \ref{druha} and paragraph
\ref{44.2} for definitions). Only then, employing the experience
gained in section \ref{druha} we shall be able to extract all the
consequences of the nontriviality of the weight operator $W\neq I$
in our Sturmian Schr\"{o}dinger eq.~(\ref{SEstcoulPT}) and its
generalizations. In particular, our main result, viz., the
spectral-series formula for the Sturmian-related metric $\Theta$
will be derived in paragraph \ref{5.}.

Finally, section \ref{summ} will offer a brief summary of our
non-Hermitian extension of the generalized eigenvalue Sturmian
problem in Quantum Mechanics where, in the terminology which
varies from author to author, the Hamiltonians $H$ and weights $W$
are admitted to be ${\cal PT}-$symmetric (in the sense of the so
called unbroken ${\cal PT}-$symmetry, \cite{Carl,BBJ}),
quasi-Hermitian (i.e., Hermitian after the inner product is
adapted, \cite{Geyer,Dieudonne}), Hermitian (in a suitable space,
\cite{Williams}) or cryptohermitian (i.e., Hermitian in a space
which we are not necessarily going to specify, \cite{Smilga}).

\section{Physics of non-Hermitian bound states \label{druha} }

Non-Hermitian Coulombic eq.~(\ref{SEracoulPT}) can be understood
as a  non-selfadjoint pair of the eigenvalue problems for $H$ and
for $H^\dagger \neq H$ with a shared real spectrum $\{\lambda\}$,
 \be
 H^{}\,|\,{\lambda}\rangle
 =\lambda\,|\,{\lambda}\rangle\,
 \label{urmjed}
 \,,\ \ \ \ \ \ \ \ \
 H^\dagger \,|\,{\lambda'}\rangle\!\rangle
 =\lambda'\,|\,{\lambda'}\rangle\!\rangle\,.
 \label{urmdva}
 \ee
The doubling of the ket symbol in some elements of our, by
assumption,  user-friendly Hilbert space ${\cal
H}^{(user-friendly)}$ merely indicates their difference at the
same real eigenvalue $\lambda$, $|\lambda\kkt \neq |\lambda\kt$.
There is nothing exotic in the latter Hilbert space since it can
be visualized as the current space $I\!\!L^2(I\!\!R)$ of the
quadratically integrable complex functions $f(x)\,\equiv\,\br x
|\,f\kt$ for which the conjugate elements of the dual space of the
functionals are obtained by the mere transposition plus complex
conjugation,  $\br f|\,x\kt \,\equiv\,f^*(x)$. At the same time,
we must keep in mind that the symbol ${\cal H}^{(user-friendly)}$
denotes an {\em unphysical} Hilbert space since we have $H^\dagger
\neq H$ in it.

In ${\cal H}^{(user-friendly)}$ we may and should assume the
validity of the standard biorthogonality, biorthonormality and
bicompleteness relations,
 \be
 \bbr \lambda'|\lambda\kt =
 \br \lambda'|\lambda\kkt =
 \delta_{\lambda'\lambda}\,,
 \ \ \ \ \ \ \ \
 I = \sum_\lambda\,|\lambda\kt\,\bbr \lambda|
  = \sum_\lambda\,|\lambda\kkt\,\br \lambda|\,.
  \label{bico}
 \ee
We can define the metric operator $\Theta=\Theta^\dagger
>0$ and ``innovate" the inner product
between the two elements $|\psi \kt$ and $|\psi'\kt$ of ${\cal
H}^{(user-friendly)}$ \cite{Geyer,BBJ,Ali},
 \be
|\psi' \kt \odot |\psi \kt
 =
 \br \psi' |\Theta |\psi \kt
 \,\equiv\,
 \bbr\psi'  |\psi \kt\,,
 \ \ \ \ \ \ \
 \Theta=\sum_\lambda\,|\lambda\kkt\,\bbr \lambda|\,.
  \label{mosta}
 \ee
The transition to this new product changes our Hilbert space  into
{\em another, unitarily non-equivalent} ``correct" and ``physical"
Hilbert space${\cal H}^{(physical)}$ of the states of the system.
Formally one could speak about an updated Hermitian conjugation
operation
 \ben
 {\cal T}^{(physical)}:|\psi \kt\,\to\, \br \psi|\,\Theta=\bbr \psi|
 \een
but its explicit use in ${\cal H}^{(physical)}$, albeit
mathematically correct, would be both unnecessary in practice and
potentially very strongly misleading.

It is clear that all the physics described in ${\cal
H}^{(physical)}$ can equally well be described in another,
unitarily equivalent, third Hilbert space  ${\cal H}^{(third)}$.
With its elements and conjugate functionals marked by the slightly
modified kets $|\chi\pkt$ and bras $\pbr \chi|$, respectively, we
reveal that we may require the unitary equivalence between ${\cal
H}^{(physical)}$ and ${\cal H}^{(third)}$,
 \be
 \bbr \chi'|\chi \kt =
 \br \chi'|\Theta|\chi\kt=
 \pbr \chi' | \chi\pkt\,.
 \ee
For this purpose it suffices that we choose {\em any} invertible
map $\Omega$ and postulate, for all the elements of the respective
vector spaces, that
 \ben
 |\psi\pkt = \Omega\,|\psi \kt\,,
 \ \ \ \ \ \
 \pbr \psi'| = \bbr \psi'|\Omega^{-1} \,.
 \een
Moreover, once we set $\Theta = \Omega^\dagger\Omega$, we may
complete our list of mappings by the formula $\bbr
\psi'|\Omega^{-1} = \br \psi'|\Omega^\dagger$. This confirms an
overall consistency of the setup where our initial Hamiltonian $H$
is non-Hermitian in ${\cal H}^{(user-friendly)}$.

In ${\cal H}^{(third)}$ the isospectral image of $H$
 \be
 \hat{h}=\Omega\,H\,\Omega^{-1}\,
 \ee
must be Hermitian, $\hat{h}=\hat{h}^\dagger$. This guarantees its
observability which is, in its turn, represented by the relation
 \be
 H^\dagger = \Theta\,H\,\Theta^{-1}\,
 \ee
i.e., in our working space ${\cal H}^{(user-friendly)}$, by the
quasi-Hermiticity of $H$.

\section{Physics of non-Hermitian Sturmians \label{tretia} }
%


In the Hamiltonian-independent Hilbert space
$I\!\!L_2(I\!\!R)\,\equiv\,{\cal H}^{(user-friendly)}$ we may
abbreviate eq.~(\ref{SEstcoulPT}) and its conjugate version as
follows,
 \be
 H^{}\,|\,{\lambda}\rangle
 =\lambda\,W\,|\,{\lambda}\rangle\,
 \label{sturmjed}
 \,,\ \ \ \ \ \ \ \ \
 H^\dagger \,|\,{\lambda'}\rangle\!\rangle
 =\lambda'\,W^\dagger\,|\,{\lambda'}\rangle\!\rangle\,
 \label{sturmdva}
 \ee
These Sturmian equations differ from their non-Sturmian
predecessors (\ref{urmjed}) solely by the presence of a nontrivial
and non-Hermitian weight $W\neq I$. Nevertheless, in a complete
parallel with section \ref{druha}  we shall still assume that all
the solutions of both of these equations are fully at our
disposal.

\subsection{Sturmian Schr\"{o}dinger equation in
${\cal H}^{(third)}$ \label{44.2}}

We intend to preserve as many analogies with the $W = I$ guide as
possible. We shall work  with a non-unitary though still
invertible mapping $\Omega$ of our space ${\cal
H}^{(user-friendly)}$ (with elements denoted by the same Dirac's
ket symbols $|\psi\kt$ as before) onto the intermediate and
abstract, third Hilbert space ${\cal H}^{(third)}$. Its elements
and their duals will be denoted by the same deformed, {\em curved}
Dirac's bra and ket symbols as above,
 \be
 |\,\psi\,\pkt = \Omega\,|\,\psi\,\kt\,\in\,{\cal H}^{(third)}\,,
 \ \ \ \ \
 \pbr\,\psi\,| = \langle\,\psi\,|\,
 \Omega^\dagger\,\in\,
 \left ({\cal H}^{(third)}\right )^\dagger\,.
 \label{lumpaci}
 \ee
The lower-case isospectral equivalent $h = \Omega\,H\,\Omega^{-1}$
of our original non-Hermitian upper-case Hamiltonians $H\neq
H^\dagger$ as well as the parallel partner $w =
\Omega\,W\,\Omega^{-1}$ of any original non-Hermitian specific
``weight" operator $W\neq W^\dagger$ are both assumed self-adjoint
in the third space.  This means that we shall require that
 \ben
 h^\dagger = \left (\Omega^{-1}
 \right )^\dagger\,H^\dagger\,\Omega^\dagger= h\,,
 \ \ \ \ \ \ \ \ \ \
 w^\dagger = \left (\Omega^{-1}
 \right )^\dagger\,W^\dagger\,\Omega^\dagger= w\,,
 \een
or, after a trivial re-arrangement,
 \be
 H^\dagger = \Theta\,H\,\Theta^{-1}\,,
 \ \ \ \ \
 W^\dagger = \Theta\,W\,\Theta^{-1}\,
 \ee
where we abbreviated $\Theta = \Omega^\dagger\,\Omega$. This is
our first result showing how the concept of the quasi-Hermiticity
translates to the Sturmian scenario.


After the above change of the Hilbert space we may represent our
upper-case Sturmian problem (\ref{sturmjed}) by its new,
lower-case reincarnation
 \be
 h^{}\,|\,{\lambda}\pkt
 =\lambda\,w\,|\,{\lambda}\pkt\,
 \label{lcsturmjed}
 \ee
which is necessarily self-adjoint  in ${\cal H}^{(third)}$. Its
simplicity facilitates the derivation of the Sturmian
orthogonality relations
 \be
 \pbr \lambda\,|\,w\,|\,\lambda'\pkt\ =\
 \pbr \lambda\,|\,w\,|\,\lambda\pkt\,\cdot\,
 \delta_{\lambda,\lambda'}\,
 \label{luma}
 \ee
and of the Sturmian completeness relations,
 \be
 I = \sum_{\lambda}\,|\,\lambda\pkt
 \,
 \frac{1}{\pbr \lambda\,|\,w\,|\,\lambda\pkt}
 \,\pbr \lambda\,|\,w\,
 \ee
as well as of the Sturmian spectral-representation formula
 \be
 h = \sum_{\lambda}\,w\,|\,\lambda\pkt\,
 \frac{\lambda}{\pbr \lambda\,|\,w\,|\,\lambda\pkt}
 \,\pbr \lambda\,|\,w\,
 \ee
for the Hamiltonian in ${\cal H}^{(third)}$.


Our original Hilbert space ${\cal H}^{(user-friendly)}$ was, by
assumption, so simple that one must always transfer all the
relevant formulae and recipes to this space at the end. In this
spirit let us insert definitions (\ref{lumpaci}) in
eq.~(\ref{luma}) and arrive at the orthogonality relations in
${\cal H}^{(physical)}$,
 \be
 \langle\,\lambda\,|\,
 \Omega^\dagger\,w\,\Omega\,|\,\lambda'\,\kt =
 \langle\,\lambda\,|\,\Theta\,W
 \,|\,\lambda'\,\kt =
 \langle\,\lambda\,|\,\Theta\,W
 \,|\,\lambda\,\kt\,\cdot\,
 \delta_{\lambda,\lambda'}\,.
 \label{deluma}
 \ee
Similarly, the appropriately adapted version of the completeness
is obtained,
 \be
 I = \sum_{\lambda}\,|\,\lambda\,\kt \,
 \frac{1}{
 \langle\,\lambda\,|\,\Theta\,W
 \,|\,\lambda\,\kt}
 \,
 \langle\,\lambda\,|\,
 \Theta\,W\,.
 \ee
Finally, the spectral decomposition of the Hamiltonian acquires
the following form in ${\cal H}^{(physical)}$,
 \be
 H = \sum_{\lambda}\,W\,|\,\lambda\,\kt
 \,
 \frac{\lambda}{
 \langle\,\lambda\,|\,\Theta\,W
 \,|\,\lambda\,\kt}
 \,\langle\,\lambda\,|\,
 \Theta\, W\,.
 \ee
This enables us to generalize the above conclusion that whenever
the spectrum of $\lambda$s remains non-degenerate, even the
generalized double-ket eigenstates $ |\,\lambda\,\rangle\!\rangle$
of $H^\dagger$ will coincide with the elementary products
$\Theta|\,\lambda\,\rangle$.


\subsection{Formula for the Sturmian metric $\Theta$ \label{5.}}

The key benefit of our pragmatic return to the space ${\cal
H}^{(user-friendly)}$ is that we may evaluate the matrix elements
$\langle\,\lambda\,|\,\Theta\,W \,|\,\lambda\,\kt$ in
eq.~(\ref{deluma}) and {\em renormalize} them to one whenever the
Hermitian product $\Theta\,W$ is positive definite (which we
assume).

This restriction will perceivably simplify our formulae. Firstly,
in terms of the other two abbreviations
 \ben
 |\,\psi\,\} = W\,|\,\psi\,\kt\,,
 \ \ \ \ \
 |\,\psi\,\ktz = W^\dagger\,|\,\psi\,\kkt\,
 \een
for elements of our working Hilbert space ${\cal
H}^{(user-friendly)}$, the two alternative forms of the simplified
orthogonality conditions will be obtained easily,
 \be
 \langle\,\lambda\,|\,\Theta\,W
 \,|\,\lambda'\,\kt=
 \brz\,\lambda\,|
 \,\lambda'\,\kt=
 \bbr\,\lambda\,|\,\lambda'\,\}=
 \delta_{\lambda,\lambda'}
 \,.
 \ee
Next we get the two alternative forms of the completeness
relations,
 \be
 I = \sum_{\lambda}\,|\,\lambda\,\kt \,
  \brz\,\lambda\,| = \sum_{\lambda}\,|\,\lambda\,\} \,
  \bbr\,\lambda\,|
 \,.
 \label{copu}
 \ee
Finally, mixed but compact expressions will result for both the
spectral-representation expansions
 \be
 W = \sum_{\lambda}\,|\,\lambda\,\} \,
  \brz\,\lambda\,|
 \,,\ \ \ \ \ \ \ \ \ \
 H = \sum_{\lambda}\,|\,\lambda\,\}
 \,
 {\lambda}
 \,\brz\,\lambda\,|\,
 \label{forumro}
 \ee
of our weight and Hamiltonian operators.

At this moment one could decide to follow the  Mostafazadeh's
$W=I$ method \cite{Alib}. Its basic idea is that one inserts the
spectral formulae (\ref{forumro}) in the quasi-Hermiticity
constraint $H^\dagger\,\Theta= \Theta\,H$,
 \be
 \sum_{\lambda}\,|\,\lambda\,\ktz \,\lambda\,\{\lambda|\,\Theta=
  \sum_{\lambda}\,\Theta\,|\,\lambda\,\} \,\lambda\,\brz
  \lambda|\,.
  \ee
This relation strongly suggests that the Sturmian analogue of the
single-series $W=I$ formula (\ref{mosta}) should be sought via
the double series ansatz,
 \be
 \Theta = \sum_{\lambda,\lambda'}\,|\,\lambda\,\ktz \,
 M_{\lambda,\lambda'}\,
 \brz\,\lambda'\,|
 \,,\ \ \ \ \ \ \ \ \ \
 M_{\lambda,\lambda'} =
 \br\,\lambda\,|\,\Theta\,|\,\lambda'\kt
 \,.
 \label{reforum}
 \ee
Fortunately, there exists an alternative approach which could have
been used, after all, also at $W=I$. Its main idea is based on the
identity $|\psi\kkt = \Theta\,|\psi\kt$ which would imply that the
Mostafazadeh's $W=I$ formula (\ref{mosta}) immediately follows
from the multiplication of the bicompleteness relation
(\ref{bico}) by $\Theta$. At $W\neq I$ the analogous idea makes us
to replace merely eq.~(\ref{bico}) by eq.~(\ref{copu}), giving the
unexpected but by far simpler, single-series final result
 \be
 \Theta=  \sum_{\lambda}\,|\,\lambda\,\kkt \,
  \brz\,\lambda\,|\,.
   \label{neebe}
 \ee
The impression of an apparent non-Hermiticity of this asymmetric
formula is misleadingn and it is virtually trivial to verify that
$\Theta=\Theta^\dagger$ in ${\cal H}^{(user-friendly)}$.

Marginally, we would like to add that as long as the alternative,
double-series formula (\ref{reforum}) for the metric is concerned,
it might still find some applications (cf., e.g., \cite{rec} for a
specific illustrative example). In similar situations our
single-series formula will still offer two useful representations
 \be
  M_{\lambda,\lambda'} =
 \br\,\lambda\,|\,\Theta\,|\,\lambda'\kt
 =
 \bbr\,\lambda\,|\,\lambda'\kt=
 \brz\,\lambda\,|W^{-1}|\,\lambda'\,\kt =
 \bbr\,\lambda\,|W^{-1}|\,\lambda'\,\}\,
  \label{reforumde}
 \ee
of the necessary nondiagonal matrix of coefficients in the less
economical but manifestly symmetric double-series expansion
(\ref{reforum}) of the Sturmian metric.

\section{Summary \label{summ} }

A return to solvable Schr\"{o}dinger equation containing Coulomb
potential helped us to establish connection between two conceptual
issues which appeared (or reappeared) in the very recent
literature on Quantum Mechanics.

The first issue reflects the well known concept of Sturmian bound
states which already found numerous and fairly diversified
applications in Quantum Mechanics over the years \cite{Rotenberg}.

The second issue involves the concept of the so called
non-Hermitian Hamiltonians with real spectra, the study of which
has been made very popular by C. Bender and his coauthors
\cite{Carl}.

An immediate inspiration of our study stemmed from two sources.
The first one was our study \cite{sgezou} where the Coulomb model
(\ref{SEracoul}) has been used as an exactly solvable illustrative
example of the survival of the reality of the energy spectrum
after a loss of manifest Hermiticity of the Hamiltonian. The
second source of inspiration can be seen in the recent new success
of an application of the Sturmina bases in computational physics
\cite{Kelbert}.

The main advantage of our present interpretation of the
observability in the non-Hermitian Coulombic as well as more
general Sturmians can be seen in its mathematical as well as
physical consistency. We made it clear that once we succeeded in
the {\em explicit} infinite-series construction of the metric
$\Theta$, even the {\em combined effect} of the ${\cal
PT}-$symmetrization and of the emergence of a nontrivial weight
$W\neq I$ still enables us to treat the resulting, {\em
apparently} non-Hermitian  Sturm-Schr\"{o}dinger equation as {\em
fully} compatible with the standard principles of Quantum
Mechanics.

In the future one could expect that the feasibility of the
construction of the Sturmian metric via formula (\ref{neebe})
could lead to its applications far beyond the present exceptional
and exactly solvable illustrative Coulomb model with ${\cal
PT}-$symmetric $W(r) \sim 1/r$. In particular, in the
topologically nontrivial class of models called ``quantum
toboggans" \cite{rec,tobo} there appeared a few new challenging
open questions which could prove tractable by our present
approach, especially (though not only) in their simplest and
exactly solvable special ``quantum knot" case of ref.~\cite{knot}
where $W(r) \sim 1/r^2$.

\subsection*{Acknowledgement}

Work supported by GA\v{C}R, grant Nr. 202/07/1307, Institutional
Research Plan AV0Z10480505 and by the M\v{S}MT ``Doppler
Institute" project Nr. LC06002.

\end{document}